\documentstyle[aps,prl,floats,graphicx,epsfig]{revtex}
\begin{document}
\draft
\twocolumn[\hsize\textwidth\columnwidth\hsize\csname @twocolumnfalse\endcsname

%
\title{Defect Formation in Quench-Cooled Superfluid Phase Transition}
\author{ V.M. Ruutu$^a$, V.B. Eltsov$^a$\cite{A1}, M. Krusius$^a$, Yu.G.
Makhlin$^a$\cite{A2}, B. Pla\c{c}ais$^b$, and G.E. Volovik$^a$\cite{A2}}

\address{ $^a$Low Temperature Laboratory, Helsinki
University of Technology, Box 2200, FIN-02015 HUT, Finland\\
$^b$Laboratoire de Physique de la Mati\`ere Condens\'ee, E.N.S., CNRS URA
1437, F-75231 Paris C\'edex 05, France}

\date{\today} \maketitle

\begin{abstract}
We use neutron absorption in rotating $^3$He-B
to heat locally a $\sim 10\;\mu$m-size volume into normal phase. When the
heated region cools
back in $\mu$secs, vortex lines are formed. We record with NMR the
number of lines as a function of superflow velocity and compare to the
Kibble-Zurek theory of vortex-loop freeze-out from a random network of
defects. The
measurements confirm  the calculated loop-size distribution and show that
also the superfluid
state itself forms as a patchwork of competing A and B phase blobs. This
explains the
A$\rightarrow$B transition in supercooled  neutron-irradiated $^3$He-A.
\end{abstract}
]

A rapid phase transition generally leads to abundant disorder and
inhomogeneity in a
heterogeneous system. But even in the ideal homogeneous case, where
extrinsic influence
is absent, a phase transition far out of equilibrium might result in the
formation of
defects. This phenomenon, if shown to be true, could explain the change
in the early universe from an initial homogeneous state to that at present
with large-scale
structure \cite{Kibble}. However, reproducible measurements on the density
and distribution of
defects as a function of transition speed are experimentally a challenging
task \cite{Zurek}.

It was recently observed \cite{BigBang2} that quantized vortices are
created in superfluid
$^3$He-B in one of the fastest 2nd order phase transitions probed to date.
The transition is
produced locally in a small volume, within the bulk medium far from
boundaries, by
irradiating $^3$He-B superflow with ionizing radiation. The most practical
heating effect is obtained from the absorption reaction of a thermal
neutron, which
creates a local overheating of suitable magnitude and volume. Vortices are
then found to
form in increasing number per absorption reaction as a function of the
superflow
velocity.

{\it Homogeneous model.}---There are several possible explanations to this
phenomenon. We show
that a quantitative comparison can be established with the theory of defect
formation in a rapidly
quenched 2nd order phase transition which was proposed by Kibble
\cite{Kibble} and Zurek
\cite{Zurek}. In this time dependent transition the order parameter of the
broken-symmetry phase
begins to form independently in different spatially disconnected regions,
by falling into the
various degenerate minima of the Ginzburg-Landau free energy functional.
Superfluid coherence is
then established only locally, in causally separated regions. These grow in
size with time and form
defects at their boundaries when they meet an adjacent region in a
different free-energy minimum.

The expected domain size of the inhomogeneity, or the characteristic length
scale in the
initial random network of defects, is $ \xi_{\rm v} = \xi_0 (\tau_{\rm Q} /
\tau_0)^{1
\over 4}$.  Here $\xi_0 \sim 20$~nm is the zero temperature superfluid
coherence
length, $\tau_0 \sim \xi_0/v_{\rm F} \sim 1$~ns the order parameter
relaxation time far
 below $T_c$, and $v_{\rm F}$ the Fermi velocity. The deviation from
equilibrium is
described by the  cooling rate $\tau_{\rm Q}$ $= [T/|dT/dt|]_{T = T_{\rm
c}}$  at  $T_{\rm c}$,
which in our $^3$He-B experiment is $\tau_{\rm Q}
\sim 5\;\mu$s. After the quench the defects relax, unless an external bias
field
is applied. In our case the superflow from rotation causes vortex
loops above a critical size to escape into the bulk liquid. There the rings
expand to rectilinear
vortex lines, which can then be counted with NMR. The bias for the
preference between $^3$He-A or
$^3$He-B can can be externally controlled with the choice of pressure or
magnetic field.

{\it Initial inhomogeneity.}---The Kibble-Zurek (KZ) mechanism has been
demonstrated to produce random networks of defects in numerical simulations
\cite{Simulation}. It
has also been compared to experiments on liquid crystals
\cite{LiquidCrystal} and superfluid
$^4$He-II \cite{Cosmology4He}. However, the KZ model describes an infinite
and spatially
homogeneous system while any laboratory sample is of finite size with
nonzero gradients.
In our $^3$He-B experiment the superfluid transition moves through the
rapidly cooling
bubble as a phase front with a width $\sim [|\nabla T|/T]^{-1}_{T = T_{\rm
c}}$.~A further important difference from the KZ model is the existence of
the broken-symmetry
phase outside the heated bubble. It might thus be the interface between the hot
bubble (with a maximum radius $R_{\rm b} \sim 30 \; \mu$m) and the bulk
superfluid
outside which governs vortex  formation.

{\it Experiment.}---The measurements are performed in a rotating nuclear
demagnetization
cryostat. The sample container is a quartz cylinder of radius $R=2.5$ mm
and height
$L=7$ mm \cite{VorNucl}. While the sample is maintained at constant
conditions, it is
irradiated with paraffin moderated neutrons from a weak Am-Be source, to
heat the fluid
locally with the nuclear reaction n + $^3_2$He $\rightarrow$ p + $^3_1$H +
764 keV. The
distance of the source from the sample is adjusted such that the observed
absorption
reactions are well separated in time. At constant neutron flux we record
the NMR
absorption as a function of time. The height of a sudden jump in the NMR
absorption
measures the number of new vortex lines which are formed in a neutron
absorption
event. Due to the large absorption cross section of the $^3_2$He nucleus,
the mean free
path of thermal neutrons in liquid $^3$He is  about 0.1 mm. Most reactions
occur
thus within a short distance from the side wall of the cylinder. Here the
superflow velocity is $v_{\rm s} =$ $\Omega R$ with respect to the wall,
when the
cylinder is rotated in the vortex-free state at an angular velocity $\Omega$
\cite{CountFlowVel}. When a vortex line is formed $v_{\rm s}$ decreases.
The reduction is
taken into account as described in Ref.~\cite{VorNucl}. In the worst case
the measuring
accuracy is $\Delta v_{\rm s} \approx \pm 0.04$~mm/s.

\begin{figure}[!!!!t]
\begin{center}
\leavevmode
\epsfig{file=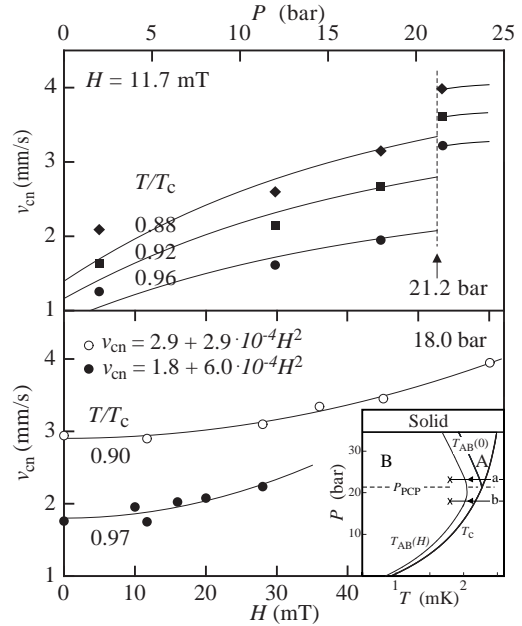,width=0.8\linewidth}
\caption[CritVel]
   {Threshold velocity  $v_{\rm cn}$ for the onset of vortex formation in
neutron-irradiated
$^3$He-B superflow: {\it (Top)} The pressure dependence displays a steep
change at $P_{\rm PCP}$,
although the B-phase properties do not change abruptly as function of
pressure.~At $P \! < \!
P_{\rm PCP}$, the curves represent $v_{\rm cn}=(\epsilon \kappa/4\pi R_{\rm
b}) \,
\ln{(R_{\rm b}/\xi)}$, where  $R_{\rm b} = $ $(3/2 e \pi)^{1 \over 2}$
$(E_0/C_{\rm v} T_{\rm
c})^{1 \over 3}$  $(1-T/T_{\rm c})^{-{1 \over 3}}$ is obtained from the
spherical thermal
model with all of $E_0 = 764$ keV transformed to heat. The fitted scaling
factor is
$\epsilon = 2.1$.  {\it (Bottom)} The dependence on magnetic field is
parabolic and reminiscent of the equilibrium state A$\rightarrow$B transition
$T_{\rm AB} (P,H) = T_{\rm c}(P) \, (1- \alpha H^2)$, where $\alpha (P)
\sim (0.5$ -- $10) \cdot
10^{-6}$ (mT)$^{-2}$ \protect\cite{Gould}. {\it (Inset)} Phase diagram of
$^3$He superfluids
with superfluid transition at $T_{\rm c}$, A$\rightarrow$B transition  at
$T_{\rm AB}(0)$ in
zero and at $T_{\rm AB}(H)$ in nonzero field, and two quench trajectories
{\it (a)} and {\it
(b)}.}
\label{CritVel}
\end{center}
\end{figure}

{\it Pressure dependence.}---Vortex lines are detected only if the superflow
velocity exceeds a threshold $v_{\rm cn}(T,P,H)$, which is plotted as a
function
of pressure $P$ and magnetic field $H$ in Fig.~\ref{CritVel}. Experimentally
$v_{\rm cn}$ is a well-defined quantity which estimates the
effective radius $R_{\rm b}$ of the heated bubble:~The largest vortex ring,
which fits
into the bubble, has a diameter ${\cal D} = 2 R_{\rm b}$. If the flow
exceeds $v_{\rm
s}=v_{\rm cn} \sim $ $(\kappa/2\pi{\cal D}) \ln{({\cal D}/\xi)}$, where
$\kappa = h/2m$
is the circulation quantum and $\xi(T,P) \approx \xi_0(P) \; (1-T/T_{\rm
c})^{-{1 \over
2}}$ the coherence length, a ring oriented perpendicular to the flow
expands into the bulk
liquid. The pressure dependence displays an abrupt increase at
about 21.2 bar, the pressure of the polycritical point: Above this pressure
$P_{\rm PCP}$
$^3$He-A is stable in zero field  below $T_{\rm c}$ between the normal and
B phases. The
measurements of $v_{\rm cn}$ are carried out well in the B phase, but when
the quench trajectory
crosses the stable A-phase regime, vortex formation is reduced (trajectory
{\it (a)} in the inset
of  Fig.~\ref{CritVel}).

{\it Magnetic field dependence.}---The parabolic dependence of $v_{\rm cn}$
on the applied field
supports the same conclusion. The only major influence of small fields on a
low pressure quench
trajectory (denoted with {\it (b)} in the {\it inset} of
Fig.~\ref{CritVel}) is to make $^3$He-A
stable in a narrow interval from $T_{\rm c}$ down to the first order AB
transition at $T_{\rm
AB}(P,H)$. Thus the magnetic field lowers the A-phase energy minimum with
respect to that of the
B phase and again this translates to a reduced yield of vortex lines at any
given value of the
bias $v_{\rm s}$.

{\it Consequences.}---The results in Fig.~\ref{CritVel}  contradict all
attempts to explain
$v_{\rm cn}$ in terms of a superflow instability \cite{VorNucl} at the
boundary of the heated
bubble, which is B phase, while A phase appears only in the hotter
interior. The
instability should occur at the $^3$He-B pair-breaking velocity $v_{\rm
c}(T,P) \approx v_{\rm
c0}(P) \; (1-T/T_{\rm c})^{1 \over 2}$ \cite{VorNucl} which is exceeded at
the bubble boundary,
unless vortices are generated more rapidly by other means. The  KZ
mechanism is inherently a
fast process: During rapid cooling through $T_{\rm c}$ the order parameter
may fall, in
different causally  disconnected regions, into A- or B-phase local
free-energy minima. Blobs of
size $ \xi_{\rm v}$ of A and B phase are formed, of which the former shrink
away in
ambient conditions, where only B phase is stable. However, it is known from
experiments with a
moving AB interface that the penetration of vortex lines through the phase
boundary is suppressed
\cite{ABinterface}. Thus we expect A-phase blobs to reduce the volume of
the initial
vortex network, confined within the B-phase blobs, and vortex formation is
impeded.

Fig.~\ref{CritVel} suggests two conclusions: 1) The  KZ mechanism is the
fastest process to
create defects, before  other phenomena, which we know from situations
close to equilibrium,
become effective. As suggested recently \cite{KV}, we may assume that the
KZ mechanism dominates
defect formation if the velocity of the phase front, at which it moves
through the heated bubble,
$v_{\rm T} \sim R_{\rm b}/\tau_{\rm Q} \sim 6$ m/s, is comparable to the
critical value $v_{\rm
Tc} \sim v_{\rm F} \, (\tau_0 / \tau_{\rm Q})^{1 \over 4}$.  2) In
supercooled $^3$He-A the KZ
mechanism starts with finite probability the A$\rightarrow$B transition. In
this case theinitial state  is supercooled $^3$He-A, subjected to ionizing
radiati
on. The final state  is the
stable  $^3$He-B, although the boundary condition favors $^3$He-A. The
deeper the
supercooling, the more likely it is that some B-phase blobs formed in a
quench merge to one
bubble which exceeds the critical diameter  and initiates the
A$\rightarrow$B transition,
as seen in experiments \cite{Leggett}. This explanation \cite{GEV} does not
require (or
exclude) Leggett's inverted ``baked Alaska'' temperature distribution in
the quench
\cite{Leggett,Bunkov}.

\begin{figure}[!!!t]
\begin{center}
\leavevmode
\epsfig{file=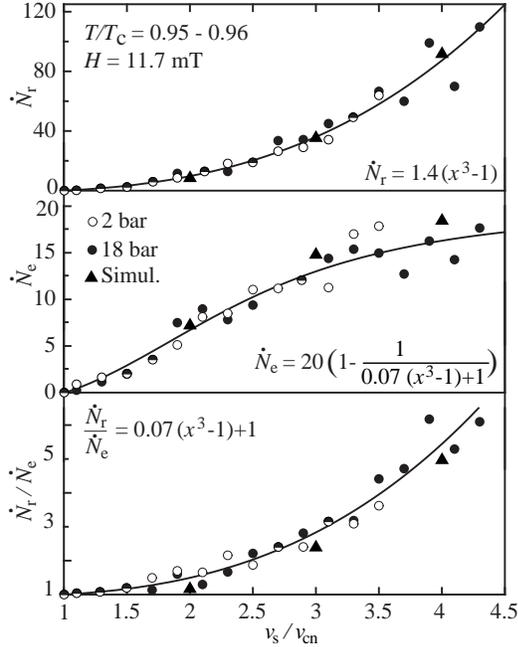,width=0.8\linewidth}
\caption[BiasDependence]
   { Rates of vortex line formation as a function of the
normalized superflow velocity $x = v_{\rm s}/v_{\rm cn}$: {\it (Top)}
Number of lines
$\dot N_{\rm r}$ and {\it (middle)} neutron absorption events $\dot N_{\rm
e}$ per
minute; {\it (bottom)} number of lines per event $( \approx \dot N_{\rm r}
/ \dot N_{\rm
e})$. All three rates have been determined {\it independently} from
discontinuities in the NMR
absorption as a function of time. The solid curves are fits to the
expressions given
in each panel. }
\label{BiasDependence}
\end{center}
\end{figure}
\begin{figure}[!!!t]
\begin{center}
\leavevmode
\epsfig{file=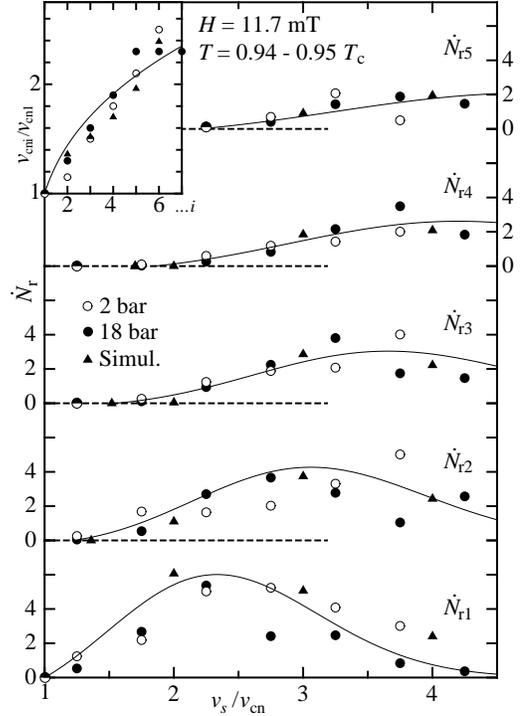,width=0.8\linewidth}
\caption[VorRingCount]
    { Rates $\dot N_{{\rm r}i}$ of vortex line formation, grouped according
to the number
of lines $i$ formed per absorption event per minute, plotted {\it vs}
$v_{\rm s}/v_{\rm
cn}$. The solid curves are guides for the eye. {\it (Inset)} Normalized
threshold
velocity $v_{{\rm cn}i}/v_{\rm cn1}$ for the onset of an event with $i$
lines, plotted
{\it vs} the number of lines $i$. The solid line represents the fit $v_{{\rm
cn}i}/v_{\rm cn1}= [2.0 \; (i-1)+1]^{1 \over 3}$.}
\label{VorRingCount}
\end{center}
\end{figure}

{\it Velocity dependence.}---Measurements of vortex-line formation as a
function of
superflow velocity $v_{\rm s}$ allow a quantitative comparison to the KZ
theory. In
Fig.~\ref{BiasDependence} we have counted per unit time the total number of
vortex lines $\dot
N_{\rm r}$ ({\it top}), the number of those neutron absorption events $\dot
N_{\rm e}$ which
produce at least one line ({\it middle}), and the number of lines extracted
from each absorption
event ({\it bottom}). The rates increase rapidly with $v_{\rm s}$: At
$v_{\rm s}/v_{\rm cn}
\approx 4.5$, close to the maximum velocity limit imposed by the
spontaneous nucleation
threshold \cite{VorNucl}, there are almost no unsuccessful (and unobserved)
absorption events
left: $\dot N_{\rm e}(\infty) - \dot N_{\rm e}(4.5 v_{\rm cn}) \approx 0$.
The data also
displays a  universality property: It can be fit to expressions like $\dot
N_{\rm r} =
\gamma[(v_{\rm s}/v_{\rm cn})^3 - 1]$, where the normalizing factor $v_{\rm
cn}(T,P,H)$ carries
all  dependence on the experimental variables. A number of tests showed  no
background
contribution in the absence of the neutron source: A vortex-free  sample
was rotated for 90 min
at different velocities (0.9, 1.3, and 2.1 rad/s at 2.0 bar and 0.94
$T_{\rm c}$), but no vortex
lines were formed.

The most detailed information is the dispersion into events in which a
given number of lines is
formed. In Fig.~\ref{VorRingCount} we plot the rates $\dot N_{{\rm r}i}$ of
events which produce
up to $i=5$ lines. This data displays large statistical variation, but after
averaging we get for each value of $i$ a curve, which is shifted to
successively higher
velocities, peaks at a maximum, and then trails off. The curves start from a
threshold velocity $v_{{\rm cn}i}$, plotted in the inset. At and
immediately above
$v_{\rm cn} = v_{{\rm cn}1}$ only single-vortex events occur. This means
that the heated
bubble resembles in shape more a sphere than a narrow cigar which is randomly
oriented with respect to the flow.

{\it Simulation.}---The initial distribution of loops in a random vortex
network with intervortex
distance $\tilde \xi = \xi_{\rm v}$ can be established with a standard
simulation calculation
\cite{VV,Simulation}.  We use a ``cubic bubble'' which is subdivided into a
grid of size
$\tilde \xi$, with up to $200^3$ vertices. A random phase is assigned to
each vertex
initially, to model the randomly inhomogeneous order parameter. On the
boundary of the
bubble the phase is fixed, to ensure that no open-ended loops are formed.
The results
have been checked by averaging over up to 1000 different initial
configurations. We use a
continuous phase variable rather than restricting it to some set of allowed
values.  The
distribution of the phase is extended to the edges of the grid according to
the shortest path on
the phase circle. Line defects are positioned to cross through the
center of those grid faces for which the phase winding is nonzero.

As usual, we assume that the initial loop distribution is preserved during
the later
evolution of the network \cite{VV}, when  the average intervortex distance
$\tilde{\xi}(t)$
increases, but the network remains scale invariant, independently of the
momentary
value of ${\tilde \xi}(t)$. Two scaling relations of standard form are
found to hold for
the networks:
\begin{equation}
  n(l) = Cl^{-\beta}~,~~~~ (C \approx 0.29,\;\;\;\;\beta \approx 2.3)~,
\label{Eq.2} \end{equation}
\begin{equation}
  {\cal D}(l) = A l^\delta~,~~~~~ (A \approx 0.93,\;\;\delta \approx
  0.47)~,
\label{Eq.3} \end{equation}
where we put $\tilde\xi=1$, $l$ and ${\cal D}$ are the length and average
straight size diameter,
and $n(l)$ the density of loops with length $l$.  The numerical values of
the parameters depend
slightly on the size of the bubble, due to boundary conditions, but in the
limit
of a large bubble they are close to those obtained for networks with mostly
open-ended strings \cite{VV,Simulation}. For a Brownian random walk in
infinite space the values
of $\beta$ and $\delta$ are $5/2$ and $1/2$.

{\it Vortex loop escape.}---The energy of a loop is
\begin{equation}
  E(l,S,t) =\rho_s \kappa ~\left[ ~l~{\kappa\over 4\pi} \; \ln
 {\tilde\xi(t) \over {\xi}}~-~ v_s S ~\right]~~,
\label{Eq.5}
\end{equation}
where $S$ is the algebraic area of the loop in the plane perpendicular to the
direction of the superflow at $v_{\rm s}$. A new result from our simulation
is a scaling law for $S$:
\begin{equation}
  |S| = B {\cal D}^{2-\nu},~~~ (B \approx 0.14,\;\;\nu \approx 0)~~.
\label{Eq.4} \end{equation}
Using Eq.(\ref{Eq.3}) for $l({\cal D})$ one has for a loop with
$S>0$
\begin{equation}
  E({\cal D},t) =\rho_s \kappa {\cal D}^2~\left[ ~{\kappa\over 4\pi
    \tilde\xi(t)A^2} \; \ln {\tilde\xi(t) \over {\xi}}~-~v_s B ~\right]~~.
\label{Eq.6}
\end{equation}
When the mean diameter of curvature $\tilde \xi(t) $ exceeds the critical
value,
$\tilde \xi_{\rm c} (v_s)= ( 1/ A^2B) \; (\kappa / 4\pi v_s) \; \ln {\tilde
\xi_{\rm c}
\over {\xi}}$, the energy becomes negative. Analytically we obtain the number
of loops extracted per neutron from $ N_{\rm r} = V_{\rm b}$ $
\int_{\tilde\xi_{\rm c}}^{2R_{\rm b}}d{\cal D}~n({\cal D})$, where
$\tilde\xi_{\rm c}(v_{\rm s}=v_{\rm cn}) = 2 R_{\rm b}$ defines the threshold
velocity. This result is only a function of the relative velocity $x = v_{\rm
s}/v_{\rm cn}$  and reproduces the measured dependence in
Fig.~\ref{BiasDependence}: $ N_{\rm  r}
\propto  x^3 - 1$.  An event with $i$ rings becomes possible, when $N_{\rm r}
\sim i$.  This gives for its threshold velocity
$v_{{\rm cn}i}/v_{\rm cn}\sim i^{1/3}$, as measured in Fig.~\ref{VorRingCount}.

In the simulation, loop  escape is modelled by setting the grid size
$\tilde\xi$ to correspond to integer values of $x=i$.  A tangled loop,
projected in the plane perpendicular to ${\bf v}_{\rm s}$, is represented as a
sum of elementary loops of grid size. The number of escaping loops $N_{{\rm
r}i}$ is assumed to be the total number of positive elementary loops. The
results in Figs.~\ref{BiasDependence} and \ref{VorRingCount} are obtained
without fitting parameters.

The scaling calculation is justified in so far that the later evolution of
the network is
orders of magnitude slower than $\tau_{\rm Q}$. Thus the latter is taken
into account
separately with a calculation of the vortex dynamics \cite{Schwarz},
including mutual friction
and the polarization of the vortex tangle by the superflow. We have
performed preliminary calculations on small lattices (up to 40 x 40 x 40)
and find that even
close to $T_{\rm c}$ the scaling law (\ref{Eq.2}) remains valid at larger
loop lengths $ l>4
{\tilde \xi}$ and that the result for $\dot N_{\rm r}(v_{\rm s}/v_{\rm
cn})$ does not change
qualitatively.

{\it Conclusion.}---We have established quantitative agreement between
measurement and the KZ
mechanism.  When the 2nd order phase transition moves at high velocity into
the volume heated by
the neutron absorption, a random network of different types of defects
is formed. This happens before other mechanisms, such as a superflow
instability at the boundary of the rapidly cooling bubble or superfluid
turbulence within its
interior, have a chance to develop. A bias field of sufficient magnitude
will select the type of
defect, which remains stable while others relax. In superflow these are
vortex loops. In
supercooled $^3$He-A it is blobs of $^3$He-B, which have  a finite
probability to start the
A$\rightarrow$B transition.

We thank Yu. Bunkov, A. Gill, Yu. Kagan, T. Kibble,  N. Kopnin, A. Leggett,
O. Lounasmaa, E.
Sonin, and E. Thuneberg. This work was funded by the EU Human Capital and
Mobility Program (no. CHGECT94-69).

\vfill\eject


\begin{references}
\bibitem{CountFlowVel} The proper variable in
Figs.~\protect\ref{BiasDependence} and
\protect\ref{VorRingCount} is the counterflow velocity $v_{\rm cf} = v_{\rm
n} - v_{\rm s}$ and
\underline{not} $v_{\rm
n} = \Omega r$
\protect\cite{VorNucl}.
\end{references}
\end{document}